\def\ital#1 { {\it #1} }
\def\JNO#1,#2,#3{{\bf#1}\ (19#2) #3}
\def\journal #1#2,#3,#4 {\ital{#1} \JNO#2,#3,#4\ }
\def\slashcap#1 {{\rlap{\kern0.25em/}#1}}

\def\vev#1 {\left\langle #1 \right\rangle}

\documentstyle[12pt,world_sci]{article}

\begin{document}

% \preprint{{\bf MITH 94/12}\\ \\ }
% \date{\today}
\title{
QUANTUM GRAVITY, THE PLANCK LATTICE\\
AND THE STANDARD MODEL
% \thanks{Invited Talk at the ``VII Marcel Grossman Meeting on
% General Relativity'' - Stanford, July 1994}
}
\author{{\bf G.~Preparata}\\
{\small\em Dipartimento di Fisica dell'Universit\`a di Milano}\\
\mbox{and}\\
{\small\em INFN, Sezione di Milano, via G.~Celoria 16, 20133 Milano, Italy}\\
{\small\em (E-mail: preparata@mi.infn.it)}
}
\maketitle

% --------------- Abstract ---------------

\begin{abstract}
A possible ground state of Quantum Gravity is Wheeler's ``space-time foam'',
which can be modeled as a ``Planck-lattice'', a space-time cubic lattice
of lattice constant $a_P\simeq 10^{33}$cm, the Planck length. I analyse
the structure of
the Standard Model defined on the Planck Lattice, in the light of the ``no-go''
theorem of Nielsen and Ninomiya, which requires an extension of the continuum
model through Nambu-Jona Lasinio terms, quadrilinear in the Fermi-fields. As
a result, a theory of masses (of both fermions and gauge bosons) is seen to
emerge that, without Higgs excitations, agrees well with observations.

\end{abstract}

% --------------- Testo ---------------

\section{Introduction}

The Standard Model (SM) of fundamental particle interactions represents the
achievement of a long scientific path that through three decades of theory
and experiment finally unifies the three different types of interactions
(strong, electromagnetic and weak) in a single and well defined theoretical
structure, that of Gauge Theories (GT).

It is remarkable that throughout the intellectual adventure that leads to the
firm establishment of the SM, Quantum Gravity (QG) is seen to play
essentially no r\^ole: the gravitational interactions, after all, at the
typical
space-time distances of particle interactions ($\sim 10^{13}$~cm)
are so incredibly tiny that the \underbar{phenomenology}
of the SM can well do without them.
Even though, and this is a tantalizing thought, the structure of QG too is
that of a GT, whose gauge group, as Einstein and his friend Marcel Grossmann
showed at the beginning of our Century, is just the Poincar\'e group.

It is only recently, after the completion of the SM synthesis, achieved at
the beginning of the Eighties with the ``announced'' discovery of the weak
gauge bosons ($W^{\pm}$, $Z^{0}$), that the widespread theoretical urge  to go
``beyond'' the SM has put the limelight on QG as one of the ``new''
interactions
that would be unified with those of the SM within the plethora of
supersymmetric
models, which have been proposed since the mid-Seventies.
Curiously enough, the various supersymmetric extensions of the SM were so
prodigal in new particles and interactions, that even the now discredited
``Fifth Force'' could easily and ``naturally'' be accommodated in their wide
and flexible framework.
However, such renewed interest in QG as a limited, particular sector of an
all-embracing, unified Theory of Everything (TOE), the main aim of
Superstring Theories, is quite different in character from that
line of thought which, at the periphery of particle physics, has been pursued
by J.A.~Wheeler and his school, that has been called Geometrodynamics \cite{a}.

The main focus of this latter research program, in fact, is to obtain a deeper
understanding of the structure of space-time, the arena of fundamental
particle interactions, that according to Einstein's General Relativity
(GR) is fully determined by the dynamics of the gravitational field.
But, as Wheeler correctly remarks, a \underbar{quantized} gravitational
field does acquire an independent dynamical r\^ole, i.e. it is no more
uniquely determined by the distribution of matter (the energy-momentum
tensor). Thus it may well happen that space-time itself, i.e. the web of
relationships among different physical events, owing to the fundamental
quantum-field fluctuations acquires some very peculiar structure, for
instance that of a ``foam'', whose discontinuities have the size of the
Planck length $a_{P} \simeq 10^{-33}$ cm.
In Wheeler's view, differently from what is being contemplated in
Supersymmetric and String Theories, QG is not a sector of an extension
of the SM, but rather the dynamical theory of space-time upon which the matter
and fields of the SM stand and evolve.

It is just this far reaching view that I will follow in this talk with the
goal to understand if and how the SM can be formulated upon a hypothesized
peculiar ``foam'' structure, that can be modeled by a simple discrete
space-time Planck Lattice (PL), whose lattice constant is $a_{P}$.
This line of research has been pursued in the last three years in
collaboration with the bright chinese theorist She-Sheng Xue.

\section{SM: what if space-time were a Planck lattice?}

Let me first remind you very briefly what is the SM on a continuous,
Minkowskian space-time (CSM).
Its Lagrangian is
\begin{equation}
\label{eq21}
{\cal L}_{CSM} (x) =
{\cal L}_{G} (x)  +
{\cal L}_{Higgs} (x)  ,
\end{equation}
where the gauge lagrangian is
\begin{equation}
\label{eq22}
{\cal L}_{G} (x) =
- {1\over4} \left(
\sum_{a=1}^{8} C^a_{\mu\nu} C^{a \mu\nu} +
\sum_{i=1}^{3} A^i_{\mu\nu} A^{i \mu\nu} +
B_{\mu\nu} B^{\mu\nu} \right) +
\sum_{q} \bar{\Psi}_{q} i \slashcap{D} \Psi_{q}  +
\sum_{l} \bar{\Psi}_{l} i \slashcap{D} \Psi_{l} ,
\end{equation}
and describes the fundamental matter fields $\Psi_{q}$ (quarks) and $\Psi_{l}$
(leptons) coupled to the gauge fields of the colour $SU(3)$-group
($C^a_{\mu\nu}$), the $SU(2)_L$-group ($A^i_{\mu\nu}$) and the
$U(1)_{Y}$-group ($B_{\mu\nu}$):
\begin{equation}
\label{eq23}
\slashcap{D} = \slashcap{\partial} + i g_1 \slashcap{B}
\left\{ {1\over2} \left( 1-\gamma_5 \right) {Y_L\over2} +
{1\over2} \left( 1+\gamma_5 \right) {Y_R\over2} \right\}  +
i g_2 {\slashcap{A} }^i {\tau^i\over2} (1-\gamma_5)
+ i g_3 {\slashcap{C} }^a {\lambda_a\over2}  ,
\end{equation}
with
\begin{eqnarray}
\label{eq24}
B_{\mu\nu} = && \partial_{\mu} B_{\nu} - \partial_{\nu} B_{\mu} , \\
A_{\mu\nu} = && \partial_{\mu} A_{\nu} - \partial_{\nu} A_{\mu}
+ i g_2 \left[ A_{\mu} , A_{\nu} \right]  , \\
C_{\mu\nu} = && \partial_{\mu} C_{\nu} - \partial_{\nu} C_{\mu}
+ i g_3 \left[ C_{\mu} , C_{\nu} \right]  .
\end{eqnarray}
All this is of course very beautiful and elegant, giving blood and flesh
to the simple and very general idea that {\bf all fundamental
symmetries must be defined locally}, thus tying intimately
symmetries to space-time.

As for the Higgs Lagrangian, its simplest form is
\begin{equation}
\label{eq25}
{\cal L}_{Higgs} (x) =
- (D^{\mu} \phi )^{\dagger} (D_{\mu} \phi) - V_{H} (\phi^{\dagger} \phi) ,
\end{equation}
with
\begin{displaymath}
\phi =
\left(
\begin{array}{c}
\phi_1 \\
\phi_2
\end{array}
\right)
\end{displaymath}
a weak isospin doublet,
\begin{equation}
\label{eq26}
D_{\mu} =
\partial_{\mu} + i g_1 B_{\mu} {Y_{H}\over2}
+ i g_2 A_{\mu}^{k} {\tau_{k}\over2}
\end{equation}
and
\begin{equation}
\label{eq27}
V_{H} (\phi^{\dagger} \phi) =
- \mu^2 \phi^{\dagger} \phi + \lambda (\phi^{\dagger} \phi)^2 ,
\end{equation}
with the ``ad hoc'' negative mass squared term to ensure ``spontaneous
electroweak symmetry breaking''.

And this is of course very ugly for:
\begin{enumerate}

\item[(a)] the elegance of the gauge principle, as applied to the fundamental
building blocks of matter leads to a {\bf massless world} and in CSM no
dynamical mechanism has been found that, respecting the Ockam's razor,
would produce the observed {\bf massive world} by using the two fundamental
building blocks only;

\item[(b)] to generate mass no better way has been found than
{\bf graft}
upon the beautifully simple gauge lagrangian ${\cal L}_{G}$ (\ref{eq22})
the ugly Higgs-mechanism, induced by ${\cal L}_{Higgs}$ (\ref{eq25}), that

\begin{enumerate}

\item[(i)] is motivated by an unpretentious pedagogical model, developed
with the aim to impressionistically describe the main features of the
Meissner effect of superconductivity;

\item[(ii)] extends in a totally arbitrary fashion the fundamental
building blocks of nature, spin-1/2 fermions and
vector bosons, to include a very odd scalar field $\phi$;

\item[(iii)] introduces the instability that leads to spontaneous
chiral symmetry breaking by means of a totally arbitrary ``negative mass
squared'' in $V_{H}$ (\ref{eq27});

\item[(iv)] introduces scalar local fields that in QFT are very peculiar
objects, there being no way, besides ``tuning'', to keep its mass from
diverging like the ultraviolet cut-off $\Lambda_{UV}$.
This is one of the main reasons why Supersymmetry, that could cure
such disease, survived that remarkable ``physics fasting'' that goes on
since more than twenty years.

\end{enumerate}

\item[(c)] the regularization/renormalization program of the QFT
defined by ${\cal L}_{CSM}$ is in bad shape, for

\begin{enumerate}

\item[(i)] the lattice regularization is blocked by the Nielsen-Ninomiya
``no-go'' theorem (see below);

\item[(ii)] dimensional regularization is still incapable of
metabolizing no less than $\gamma_5$!

\end{enumerate}

\end{enumerate}

In spite of all these difficulties the low energy phenomenology of
electroweak interactions is fundamentally decoupled from all such flaws,
that are all of a conceptual, theoretical nature, and stands as a beautiful
confirmation of the simple and powerful $SU(2)_{L} \otimes U(1)_{Y}$
gauge principle.
And this is the strong message of 20 years of electroweak interactions.

In view of all this, the question of the title of this Section appears
much less philosophical and may well lead us to a SM without all the defects
that have been exposed above, as I shall try to argue in the rest of this
Lecture.
As mentioned in the Introduction, I side completely with John Wheeler
who sees the violent quantum fluctuations at the Planck scale to tear
continuous space-time apart and create a foamlike structure, full of voids and
discontinuities\footnote{The analogy with the non-trivial vacuum structure
that has been demonstrated \cite{r1} to emerge in another non-abelian GT, QCD,
is to my mind particularly relevant.
I hope to be able to make it more precise in the near future}.

Thus our question can be reformulated as: what if Wheeler's
idea were right and space-time at distances smaller than $a_{P}$ would
dissolve into the nothingness of wormholes?. A positive answer would
 lead us to make the following hypothesis:
space-time is {\bf not} a 4-dimensional continuum but rather
a (random) Planck lattice.

As a consequence the SM and all QFT's, that so far have been defined
on a continuous manifold, must be reformulated  as Lattice QFT's.
It is amusing to note that  our hypothesis, if right, would  turn the various
``theories of everything'' into theories of nothing, for at the scales where
superstrings become relevant space-time would dissolve. But if we are to
formulate the SM as a chiral Lattice Gauge Theory (LGT), our research program
clashes immediately with the "no-go" theorem of Nielsen and Ninomiya \cite
{r2},
which informs us that we cannot simply transcribe the usual SM on a lattice,
for when a LGT is chiral the low energy spectrum (the massless fermions)
gets doubled in such a way that the fundamental chiral symmetry is violated.
The physical origin of such  unpleasant result is the peculiarity of the
dispersion relations in a discrete space-time: for a Weyl fermion
$[ \Psi_L={1\over 2}(1-\gamma_5)\Psi]$ the free-field dispersion relation is
\begin{equation}
\label{eq28}
\omega(\vec p)={1\over a}\sum_i \sin(\vec p_i a),
\end{equation}
yielding low-energy ($\ll 1/a$) solutions not only for $\sum_i\vec p_i
a \simeq 0$, but also for
$\sum_i\vec p_i a \simeq \pi$. And the existence of these ``doublers" is
enough, not only to create problems with the observed spectrum, but also
to destroy the very notion of left-handedness that permeates the
electroweak phenomenology. Indeed it is easy to show that``doublers" do not
have
the same chirality of the Weyl field. Thus the ``no-go" theorem of Nielsen and
Ninomiya would appear to bar as velleitarian and impossible the idea
of a Planck Lattice Standard Model (PLSM), whose foundation is the object
of this talk.
However, a careful analysis of its proof turns out to suggest very
clearly the necessary ingredients towards a viable  formulation of PLSM.
Indeed,
one of the crucial hypotheses of the ``no-go" theorem is that the SM action be
{\bf bilinear} in the matter fermionic fields. As well known, this
is precisely the structure of the usual Wilson action \cite{r3} ($P_{L,R}=
1/2(1\pm\gamma_5)$, F is a fermion index)
\begin{equation}
\label{eq29}
S_D^F =
{i\over 2}\sum_{x,\mu,F} \left\{
\bar\Psi^F(x)\gamma_\mu U_\mu(x) [L_\mu^F(x)P_L+
R_\mu^F(x)P_R]\Psi^F(x+u_\mu) - \ital{h.c.} \right\} ,
\end{equation}
where $U_\mu (x) \in SU(3)$ and $L_\mu (x) \in SU(2)_{L}\otimes U(1)$ and
$R_{\mu} \in U(1)_Y$.

Thus if one violates this hypothesis by introducing the simplest
generalization of the SM action, i.e. adding interaction terms that are
{\bf quadrilinear} in the Fermi fields, one may have a chance to overcome
this unpleasant obstacle.
Furthermore, this addition would introduce into the SM a term of the Fermi
type, that was proposed more than 30 years ago by Y.~Nambu and
G.~Jona-Lasinio \cite{r4}, in the context of chiral symmetry breaking
in QFT!
Also note that quadrilinear terms of the type we will consider are
expected to arise as ``effective'' interactions  among matter fields due to
the gravitational interactions at the Planck scale, where gravitation ceases
to be so utterly negligible.

Thus the PLSM action is:
\begin{equation}
\label{eq210}
S_{PL} = S_{G} + \sum_{F} \left( S_D^F + S^F_{NJL1} + S^F_{NJL2} \right)  ,
\end{equation}
where ($G_{1,2}$ are $O(a^2)$, F denotes the fermions -- leptons and quarks
-- $i$, $j$ denote the $  SU(2) \times U(1)$ indices)
$S_G$ is the usual Wilson gauge lagrangian \cite{r3}, $S_D$ the
bilinear Dirac lagrangian, and
\begin{equation}
\label{eq211}
S^F_{NJL1} = - G_1 \sum_{x}
\bar{\Psi}_L^F (x)^i \Psi_R^F (x)_j
\bar{\Psi}_R^F (x)^j \Psi_L^F (x)_i ,
\end{equation}
\begin{equation}
\label{eq212}
S^F_{NJL2} = - {G_2\over 2} \sum_{x \mu}
\bar{\Psi}_L^F (x)^i
L^F_{\mu} (x)^{i'}_{i} U^F_{\mu} (x)
\Psi_R^F (x+a_{\mu})_{j'}
\bar{\Psi}_R^F (x)^j
R^F_{\mu} (x)^{j'}_{j} U^F_{\mu} (x)
\Psi_L^F (x+a_{\mu})_{i'}.
\end{equation}
Complicated though they appear, $S^F_{NJL1,2}$ simply enforce the
$SU(3)_c \otimes SU(2)_L \otimes U(1)_Y$ gauge-principle and evade the
``no-go'' theorem, at least in principle, {\bf without the
introduction of new matter or gauge fields}.
As noted above, I wish to emphasize once more that terms like these are indeed
expected to arise as effective matter interactions, induced by the
gravitational interactions at the Planck scale.
Their size too $[O(a_P^2)]$ is just what one expects from interactions of
gravitational origin.

The question now is: does the action $S_{PL}$
(Eqs.~(\ref{eq210})-(\ref{eq212})) give rise to a SM consistent with the
multiform and sophisticated ``low-energy'' phenomenology?
The next Section is devoted to giving a convincing, positive answer to this
most important question.

\section{A Consistent SM on the Planck lattice}

As I have just stressed, our $S_{PL}$ action evades in principle the ``no-go''
theorem, will it also evade it in practice?
This is the question on which She-Sheng and myself are devoting quite
a lot of hard thinking since three years, and this is where we have got so far
\cite{r5}

\subsection*{(a) The origin of masses}

We start our analysis by computing the energy of the ground state (the
effective potential) as a function of (we work from now on with a euclidean
lattice; $V_4$ is the 4-dimensional  euclidean volume)
\begin{equation}
\label{eq31}
M^{F} = - {G_1\over 2 V_4} \sum_{x}
\vev{\bar{\Psi}^{F} (x) \Psi^{F} (x)},
\end{equation}
and
\begin{equation}
\label{eq32}
r^{F} = - {G_2\over 8 V_4} \sum_{x , \mu}
\vev{\bar{\Psi}_{L}^{F} (x) U_{\mu} (x) \Psi_{R}^{F} (x+a^{\mu}) +
\mathrm{h.c.} } .
\end{equation}
Note that both $M^F$ and $r^F$, when different from zero, violate the
original chiral gauge symmetry, causing some or all the fermions to acquire a
mass and {\bf all} the doublers to ``go to Heaven'', i.e. to get a mass
of the order of the Planck mass $m_{P} \simeq 10^{19}$ GeV.

In other words, what we are trying to ascertain is whether, by generating
non-zero values of $M^F$ and $r^F$, the ground state energy gets lowered with
respect to the symmetric situation where $M^F = r^F = 0$.
Note that the symmetric situation corresponds to a totally unacceptable,
because unrealistic, scenario where all the fermions and gauge bosons are
massless, the low-energy spectrum is doubled and no evasion from the
``no-go'' theorem is possible.

The problem we must now solve is formally identical to the one that occurs
in the theory of Superconductivity, where a key r\^ole is played by the
possible non-trivial solution(s) of the gap-equation(s).

To have a clearer idea of what kind of physics is involved, as a first step
we switch off the gauge-interactions and keep the Nambu-Jona Lasinio (NJL)
terms only.
In the mean-field approximation (the same that is effectively employed
in the theory of Superconductivity) it is a completely straightforward
business to derive the appropriate ``gap equations'' and to solve them with the
following results:
\begin{enumerate}
\item[(i)] $r^F$ is non-vanishing for {\bf all} fermions and for all
choices of the adimensional coupling constants $g_2 = {G_2\over a_P^2}$
(Fig.~1). This result fulfills the necessary and sufficient condition for the
removal of doublers;

\begin{figure}[hbt]
\begin{center}
\leavevmode
\hbox{ }
\vspace{10cm}
\hbox{ }
\caption{The function $r^F(g_2)$ for quarks (q) and leptons(l) ($ma\simeq 0$).}
\end{center}
\end{figure}

\item[(ii)] $M^F$ -- the fermion mass matrix -- also turns out to be in
general different from zero, depending however on the values of
$g_1$ and $g_2$ (see Fig.~2).
Many solutions, however, are possible with the following features:
\begin{enumerate}
\item[($\alpha$)] all leptons are massless;
\item[($\beta$)] either one, two or three quark families become massive
(the other two, one or zero remaining massless).
\end{enumerate}
\end{enumerate}

The real minimum of the ground state energy (the effective potential) can
be ascertained, and the true solution can be found, only when the
problem of composite particles (the Goldstone modes) is solved and their
contribution to the ground-state energy is computed.
This will be discussed in a moment, for the time being, however,
we remark that:
\begin{enumerate}
\item[(i)] mass is generated in the SM;
\item[(ii)] doublers are removed;
\end{enumerate}
and all this without arbitrarily extending the basic building blocks (fermionic
matter and
gauge-fields) of the SM.

\begin{figure}[hbt]
\begin{center}
\leavevmode
\hbox{ }
\vspace{10cm}
\hbox{ }
\caption{The critical line for $m=0$ in the $g_1-g_2$ plane.}
\end{center}
\end{figure}

\subsection*{(b) The composite particles' spectrum}

When one analyzes the correlation functions

\vskip 1cm

\begin{equation}
\label{eq33}
\hskip3cm= - {G_1\over\sqrt{2}}
\vev{\bar{\Psi}^{F} (x) \gamma_5 \Psi^{F}(x)
\bar{\Psi}^{F} (0) \gamma_5 \Psi^F(0) } ,
\end{equation}

\vskip 1cm

one finds, as predicted by the Goldstone theorem, massless poles
-- the Goldstone bosons --
whose number is equal to $(2 N_F)^2$, where $N_F$ is the number of quark
families that acquire a non-zero mass.

Note that the solution $N_F \not= 1$ is phenomenologically disastrous,
for only four Goldstone bosons can be ``incorporated'' in the
longitudinal dynamics of the gauge fields, which become thus massive,
as experimentally observed.
All Goldstone bosons in excess of four should therefore remain in the
observable
spectrum, thus clashing with the experimental fact that no such particles
have ever been detected, even though they could be copiously produced.

In addition one can analyze the scalar correlation function

\vskip 1cm

\begin{equation}
\label{eq33a}
{G_1\over\sqrt{2}}
\vev{\bar{\Psi}^{F} (x) \Psi^{F}(x) \bar{\Psi}^{F} (0) \Psi^F(0)}  =
\hbox to 4cm {\hfil}
\end{equation}

\vskip 1cm

and look for possible poles in this channel too.
This exercise is far from academic, for in the continuum SM supplemented
by NJL-interactions, studied in refs.~\cite{r6} , it was found
that the ugly and unnatural Higgs particle, chased away from the
theory and replaced by the NJL-term, slyly reappears in the particle spectrum
as an ``almost pointlike'' composite scalar particle with mass
\begin{equation}
\label{eq35}
m_{S} \simeq 2 m_{F} = 2 m_{top} ,
\end{equation}
a very high mass, which according to current knowledge should be around
400 GeV: a very natural goal of the future (if any) supercolliders.
The remarkable and surprising consequence of the novel structure of the PLSM
at the Planck
scale is that instead of Eq.~(\ref{eq35}), we obtain
\begin{equation}
\label{eq36}
m_S^2 = (2 m_F)^2 + 0.8 m_F {r_F\over a_P} + 0.9 {(2 r_F)^2\over a_P^2} ,
\end{equation}
where $r_F \simeq 0.25$ is the coefficient of the Wilson-term, that is
responsible for the lifting of the doublers' mass to the Planck scale.
Thus, according to (\ref{eq36}), the finite mass (\ref{eq35}) of the CSM
receives in the PLSM corrections of the order of the Planck mass itself
$1/a_P = m_{P}$.
The physical origin of the extra terms that lift the Higgs mass to $m_P$
is quite easy to track: it stems from the doublers that, due to
$r^F \not= 0$, dominate the spectrum at the Planck scale.

\section{The origin of masses}

Having found a consistent way to build a PLSM, by evading the constraints of
the "no-go theorem" through a minimal set of NJL quadrilinear terms, we shall
now investigate how such PLSM takes account of the most fundamental and
puzzling
aspect of the electroweak phenomenology, that of fermion (quarks and leptons)
and boson masses.

In the preceding Section we have seen that in the simple mean-field
approximation to the ground state energy density, the gap equations for the
fermion masses allow a plurality of solutions, where one, two or three quark
generations acquire a non-zero mass, while the rest of fermions, and in
particular {\bf all} leptons, remain massless. There exist also
solutions where the situation between quarks and leptons gets interchanged,
depending on the (very) fine-tuning of the coupling constants $g_{1,2}$.
It is interesting that such asymmetry between quarks and leptons finds its
origin in the colour degree of freedom, which the leptons lack.

For obvious phenomenological reasons we must choose the former (exclusive)
class of solutions, and in order to see whether the degeneracy noted above
may be lifted
we must go beyond the mean-field approximation to the vacuum energy, to
include the contributions from the composite fields, that we have just
seen to emerge from the solutions of the gap equations. By adding such
contributions to the effective potential we obtain the energy density
$\Delta E(r)$ ($\Delta E=0$ for the symmetric configuration) as a function of
$r$ (the Wilson parameter) for $m^F a_p\simeq 0$ depicted in Fig.~3.

\begin{figure}[hbt]
\begin{center}
\leavevmode
\hbox{ }
\vspace{10cm}
\hbox{ }
\caption{The vacuum energy $\Delta E(r_q)$ for different $N_F$
($m^qa_P\simeq 0$).}
\end{center}
\end{figure}

Remarkably, only the solution with $N_F=1$ realizes an energy gain with
respect to the symmetric ground state, attaining its minimum at $r\simeq 0.25$.
The other two solutions of the gap equations with $N_F=2$ and 3 are always more
energetic than the symmetric, massless solutions. This results is extremely
 welcome, for it implies that in a first stage of approximation the fermionic
spectrum contains only two massive quarks (the t- and b-quarks), while all
other
fermions are massless. Furthermore, as we shall see in a moment, the four
Goldstone bosons that, in the absence of gauge-interactions, result from
the"spontaneous" breaking of the original chiral symmetry, are in the right
number and structure to get incorporated in the longitudinal dynamics of the
weak gauge bosons, thus giving rise to the observed masses of such particles.

When we switch the gauge interactions on, and analyse the Dyson's equations of
 the PLSM for the fermion self energies we obtain for the self-mass operator
 of the massive doublet (top and bottom) the equation depicted in Fig.~4.

\begin{figure}[hbt]
\begin{center}
\leavevmode
\hbox{ }
\vspace{6cm}
\hbox{ }
\caption{The Dyson equations for top and bottom quark masses.}
\end{center}
\end{figure}

What is so interesting about such equation,when evaluated on the Planck
lattice, is that, due to the doublers, at the Planck scale the charged gauge
bosons ($W^{\pm}$) contribute a term which couples $\Sigma_t$ to
$\Sigma_b$ and renders the above equation equation non diagonal. Furthermore
the different e.m. charges of the top- (2/3) and bottom- (-1/3) quark
introduce an asymmetry capable in principle to lift the degeneracy between the
two
 quarks, that in a pure NJL-scenario are indeed degenerate. But, can it lift
it in the dramatic way suggested by observation? A completely straightforward
calculation \cite{r5} gives the very enticing answer:

\begin{equation}
\label{41}
 m_t\simeq 30 m_b,
\end{equation}

which setting $m_b \simeq 5 GeV$, predicts the top mass in the ball park where
it has been very recently announced.
Another very pleasing aspect of the PLSM analysis, as compared with the one
carried out within the CSM \cite{r6}, is that the tuning of the coupling
constant
$g_{1,2}$, necessary to yield finite quark masses, is not so incredibly fine,
i.e. of O$(m_q/m_P)$, but only of order O$(\alpha\log(m_q/m_P))$.
The same exercise can be carried out for
other quarks and leptons, where now the composite fields' contribution is
missing, and for leptons where the NJL-term is 3 times smaller than for the
quarks, due
to their being colourless. When it is done (but it has not yet been completed)
a complete fermion spectrum should emerge tying its apparently capricious
pattern to the mixing (Kobayashi-Maskawa) matrices of the quarks and leptons.
The potentiality of this approach looks really remarkable: I hope to be able to
report its final results soon.

Turning now to the masses of the gauge bosons, in order to calculate them
we must analyse the gauge-bosons self energy function.

\begin{equation}
\label{4.2}
\Pi_{\mu\nu}(k)=(-g_{\mu\nu}+{k_\mu k_\nu \over k^2})\Pi (k^2)
\end{equation}
given by the diagrams (Fig.~5).

\begin{figure}[hbt]
\begin{center}
\leavevmode
\hbox{ }
\vspace{7cm}
\hbox{ }
\caption{The diagrammatic form of the Dyson equation for $\Pi_{\mu\nu}(k)$.}
\end{center}
\end{figure}

Doing the tedious job of computing the diagrams \cite{r5}, and in particular
keeping
track of the composite Goldstone modes, we obtain a relation between the
top-quark mass $m_t$ and the $Z$-mass:
\begin{equation}
\label{43}
m_t=1.633 M_Z=149 GeV.
\end{equation}

Furthermore we obtain:
\begin{equation}
\label{44}
{M_W^2\over M_Z^2}=\cos^2\Theta_W(1+\rho),
\end{equation}

where
\begin{equation}
\label{45}
\rho=({m_b\over m_t})^2 {\log({M_Z^2\over M_W^2})\over \log({\Lambda_P^2
\over m_t^2})}\simeq 10^{-6},
\end{equation}
in agreement with well established phenomenology.

\section{Conclusions}

Let me now conclude this talk by trying to summarize and put in perspective
its theme. As I have already emphasized in the Introduction, in its
momentous development since the end of World War II, that led to the
admirable synthesis of the SM, Particle Physics rarely felt any need
or interest for Gravitation, the particle aspect of which, all the way down
to the Planck scale, looking so incredibly remote. And if recently
the attitude towards Gravitation has changed, it has only been in the framework
of a far fetched search for the Theory of Everything, a definite act of
hybris which may well ruin theoretical particle physics.

Fortunately in Astrophysics and Cosmology the great tradition that from
Riemann through M. Grossmann and A. Einstein enriched mankind with the
classical theory
of Gravitation (General Relativity), has been kept alive and keenly focused
on the fundamental problem of the structure of Space-Time, as determined by
the dynamics of Gravity: Geometrodymanics, as J.A.~Wheeler has aptly
called it. It is this way of looking at Gravity, within the general framework
of QFT, that is giving new meaning and relevance to the "neglected
interaction" of Particle Physics, which now plays the fundamental role of
determining the very nature of the space-time, in which the drama of the SM is
played.
I believe it will soon be possible to show that, as conjectured by Wheeler
himself \cite{a}, at the Planck scale Euclidean Space-Time - the ground state
of classical Gravity - will cease to be the ground state of Quantum Gravity
and it will be replaced by a "condensate" of "wormholes", of
size $a_P$ at a distance $a_P$: a foam of Planck size. The reason for
believing in the soundness of Wheeler's intuition is a similar result
that I have been able to derive for the (likely) ground state
of another non-abelian gauge-theory: QCD \cite{r1}. There, the highly
non-linear
interactions of the quantum fluctuations -- the gluons --  render a finely
structured network of needled shaped chromomagnetic domains highly stable
with respect to the perturbative ground state; and as a consequence on such
a ground state  all isolated colour charges acquire an infinite mass,
thus disappearing from the physical spectrum: colour confinement is finally
explained and understood.

And if space-time at the Planck scale dissolves into a foam of wormholes,
it is a fascinating question to ask what will then happen to the local matter
and gauge fields of the Continuum Standard Model (CSM). In such a discontinuous
structure, which one can reasonably approximate with the more regular
structure of a Planck lattice, She Sheng Xue and I have gone through the long
and to try give an answer to this questionr. The immediate difficulty posed
by the "no-go" theorem, that bans the
possibility to directly transcribe the CSM on any lattice surprisingly
\underbar{unveils}
a path -- the addition of a minimal number of quadrilinear Fermi interactions
(see Eqs. (2.13) and (2.14)) -- that allows us to get rid of the ugly and
unnatural Higgs
Lagrangian (Eq. (2.7)). In fact the fundamental problem of the generation of
the masses
of both fermions and weak gauge bosons gets an immediate solution
through a process of condensation of fermion-pairs, quite analogous
to what happens in the phenomena of superconductivity. Mass does get generated
\underline{ in principle} without adding to the beautiful and
economical SM an artificial and troublesome new type of matter, the
Higgs scalars. Do the generated masses conform with the seemingly chaotic
observed pattern? Even though the necessary analysis is not yet completed,
we can  however state with confidence that in the "spontaneous" symmetry
breaking
mechanism of the PLSM:

\begin{enumerate}
\item[(i)] a single quark family -- the t,b-family -- receives masses that are
much larger than those of the other quark and lepton families;
\item[(ii)] the $m_t/m_b$-ratio is predicted correctly [Eq.~ 21];
\item[(iii)] a correct relation can be derived between the top mass and
the weak boson masses [Eqs.~(23), (24) and (25)]
\item[(iv)] a theory of the mixing matrices for both quarks and leptons seems
within reach.
\end{enumerate}

If the promise will finally be kept and a successful PLSM will be added to
the treasures of scientific rationalism, a new and intriguing type of
relationship will be seen to tie Gravity to the masses of the Universe: no more
 and only the sources of Gravity, masses will finally have found their
``raison d' \^etre" in the violent fluctuations of the metric gravitational
field that at the Planck distance tear space-time apart.

% --------------- References ---------------

\def\these {these Proceedings.}
\def\etal  {\ital{et al.}}


\begin{thebibliography}{00}

\bibitem{a} C.W. Misner, K.S. Thorne and J.A. Wheeler, Gravitation
(Freeman, San Francisco 1973).


\bibitem{r1} See for instance G.~Preparata,
\ital{Why Are Quark (and Gluons) Confined ?}, MITH 93/18, Milano, June 1993,
Preprint, to appear in the Proceedings of the VIII Winter School on
``Hadronic Physics'' -- Folgaria, 1-6 February 1993.

\bibitem{r2} H.B. Nielsen and M. Ninomya, Nucl. Phys. B185(81)20 and
B193(81)173, Phys. Lett. B105 (1981) 219.


\bibitem{r3} K. Wilson, in "New phenomena in subnuclear physics" (Erice, 1975)
ed. A. Zichichi (Plenum, N.Y. 1977).

\bibitem{r4} Y. Nambu and G. Jona-Lasinio, Phys. Rev. 122(1961)345.

\bibitem{r5} G. Preparata and S.-S. Xue, Phys. Lett. B264(91)35; B302(93)442;
B235(94)161; B329(94)87; B335(94)192; B325(94)161;``Emergence of the $ t\bar t$
condensate
and the disappearance of Higgs scalars in the Standard Model on the Planck
Lattice", preprint MITH 93/5, submitted to Nucl. Phys.

\bibitem{r6} W.A. Bardeen, C.T. Hill and M. Linder, Phys. Rev. D41(90)1647.

\end{thebibliography}
\end{document}